\documentclass[pre,twocolumn,twoside,latterpaper,floatfix]{revtex4} 
\usepackage{graphicx}   
\usepackage{epsfig}
\usepackage{latexsym}

\begin{document} 
\title{Do 
 methanethiol adsorbates on the Au(111) surface dissociate?} 
\author{Jian-Ge Zhou, Frank Hagelberg}   
\affiliation{Computational Center for Molecular Structure and Interactions, 
Department of Physics, Atmospheric Sciences, and General Science, Jackson 
State University, Jackson, MS 39217, USA }  
\author{}   
\affiliation{}  
  
\begin{abstract}
The interaction of methanethiol molecules CH$_{3}$SH with the Au(111) surface is investigated,
and it is found for the first time that the S-H bond remains intact when
the methanethiol molecules are adsorbed on the regular Au(111) surface. 
However, it breaks if defects are present in the Au(111) surface. 
At low coverage, the fcc region is favored for S atom adsorption, but
at saturated coverage the adsorption energies at various sites are almost iso-energetic. 
The presented calculations show that a methanethiol layer on the regular Au(111) surface does not dimerize.
\end{abstract}  
\maketitle

Self-assembled monolayers (SAM) are thin organic films that form spontaneously on solid surfaces.
These systems have been the subject of intense research in recent years, both experimental and computational, 
see Ref. \cite{au}
for a review.
This level of interest may be ascribed to their importance in wetting phenomena, tribology, chemical and biological sensing,
optics and nanotechnology. Among the many varieties of SAM, the adsorption process of alkanethiol molecules
on the Au(111) surface has been given special attention because of the relative simplicity of the molecules,
the highly stable and ordered SAM structures, and the ease of preparing  the Au(111) surface.
Despite the apparent simplicity of this system, its observation in various experiments has led to controversial results. 
One of the key issues is whether S-H bond dissociation might occur for the alkanethiol molecule 
when adsorbed on Au(111) surface. 

Numerous experiments have focused on the alkanethiol adsorption on the Au(111) surface.
These efforts have given rise to the long-standing controversy whether the S-H bond of methanethiol molecules 
adsorbed to the Au(111) surface is dissociated or not \cite{au}.
Nuzzo et. al first observed non-dissociative adsorption
of thiols on the surface by use of electron-energy-loss spectroscopy (EELS), X-ray photoemission and 
temperature-programmed
desorption (TPD) \cite{nzd}. Based on the shape of the potential-energy diagram scaled with the heat of alkanethiol
adsorption on Au(111), it was suggested that S-H bond dissociation may occur for alkanethiols \cite{dzn}. Kodama et al. reported thiolate radical desorption
for alkanethiol from the Au(111) surface \cite{khn}. The TPD and X-ray  photoelectron spectroscopy (XPS) studies by
Liu and co-workers indicated the dissociative adsorption of the methanethiol on Au(111) \cite{lrd}.
The pioneering theoretical work of Groenbeck and his coworkers \cite{gca} showed that
the S-H bond should be cleaved once the methanethiol molecules are adsorbed on the Au(111) surface. In the wake of this work, 
it was generally assumed that dissociation takes place, and the cleavage of the S-H bond was used as a premise for further 
modeling of the methanethiol adsorbate on Au(111) \cite{au}. However, in a recent TPD, Auger electron spectroscopy (AES),
and low-temperature scanning tunneling microscopy (LT-STM) study of this system, no scission of the S-H bond \cite{rlm} was found.
So far, no consistent and unified theoretical interpretation of the dissociation problem of methanethiol 
on the Au(111) surface has been presented. 
Since different experiments, investigating the same system from different angles, arrived at different conclusions, theory is challenged to propose a consistent model for the methanethiol adsorption process on the Au(111) substrate.

Guided by this motivation, in the present letter we propose 
a consistent and unified model to clarify the controversial S - H bond breaking issue.
First we consider the adsorption energies for the
non-dissociative and dissociative configurations in the $c(4\times 2)$ 
superlattice at low coverage (0.25ML), 
and report for the first time that the non-dissociative structure is favored for the regular
Au(111) surface. Then we study a set of eleven non-dissociative
structures at high coverage (1.0ML). Several initial
configurations are designed as sulfur dimers, but after optimization the 
dimers disappear, thus we find that for methanethiol molecules the sulfur dimerization is not favored even at 
high coverage. To identify the conditions under which dissociation can occur, we make admission for defects 
in Au(111). The S-H bond cleavage is found to proceed in the presence of 
defects in the gold surface. We thus report a new mechanism for the scission of methanethiol on Au(111).

\begin{figure}
\includegraphics[width=2.25in]{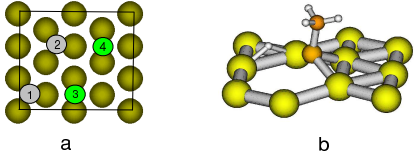}
\caption{(a) The c(4x2) supercell used in the calculation. (b) The stable dissociative
structure with the vacancy.}  
\label{4x2}
\end{figure}

Employing the VASP code \cite{khf},
we follow two avenues of computation, using 1) the PAW potential \cite{kj},\cite{peb} which involves 
the generalized gradient approximation (GGA) \cite{pw} to exchange and correlation (XC) in the framework of 
the PW91 approach; 2) the ultrasoft pseudopotential \cite{khf} with XC 
based on the 
PBE formalism \cite{pbe}.
The wave functions are expanded in a plane wave basis with an energy cutoff of 400 eV. The Brillouin zone integrations 
are performed by use of the Monkhorst-Pack scheme\cite{mp}. 
We utilized a $3\times 3\times 1$  k point mesh for the geometry optimization.
The Au(111) surface supercell consists of a $c(4\times 2)$ superlattice (see Fig. \ref{4x2}a), i.e. 
12 Au atoms per layer,
a total of 48 Au atoms. The Au atoms in the 
top three atomic layers are allowed to relax, 
while the Au atoms in the bottom layer are fixed to simulate 
bulk-like termination\cite{csb}-\cite{zhx}. 
The vacuum region comprises seven atomic layers.
We increased the energy cutoff to 500 eV and the number of 
k points to $8\times 8\times 1$. 
These two cases differ by
less than $2\%$. We calculated the gold lattice constant, 
and found it to agree with the experimental value \cite{ksu} within  $2.1\%$. 

First we comment on our results related to the non-dissociative geometries and adsorption energies for 
the optimized configurations 
of the methanethiol molecule on the Au(111) surface at 0.25ML, as displayed in Table \ref{phys}. 
The units for the bond length and adsorption energy are Angstrom ($\AA$) and eV, respectively. 
\begin{table}
\caption{The non-dissociative geometries and adsorption energies for the structures 
of methanethiol on Au(111) at 0.25ML. The entries $\theta$, $tilt$ $direct$ and $d_{S-Au}$  refer to
the polar angle between the normal vector of the surface and the S-C bond direction, the Au(111) surface region 
towards which
the S-C bond is tilted, and the shortest Au-S bond length. The entries $initial$ and $optimized$ $site$ stand for
the S atom attachment site before and after optimization. The columns 2-4 and 6-8 list structural data 
pertaining to the initial
and the final optimized geometry. The maximum adsorption energy is underlined.}
\begin{center}
\begin{tabular}{cccccccc}
\hline
initial & $\theta$ & $d_{S-Au}$ & optimized & $\theta$ & tilt     & $d_{S-Au}$ & $E_{ads}$\\
site    & ~        & ~       & site      & ~        & direct & ~       & ~\\
\hline
  ~&0&2.60&bri-fcc&11.4&fcc&3.13&0.44\\
fcc&45&2.60&bri-fcc&33.1&hcp&2.83&0.42\\
  ~&90&2.60&fcc&83.0&hcp&3.65&0.42\\
\hline
  ~&0&2.60&bri-hcp&11.6&hcp&3.01&0.40\\
hcp&45&2.60&bri-hcp&43.1&fcc&2.87&0.33\\
  ~&90&2.60&hcp&89.7&fcc&3.67&0.43\\
\hline
  ~&0&2.60&bri-hcp&4.8&fcc&2.89&0.43\\
  ~&45&2.60&bri&50.9&hcp&3.30&0.45\\
 bri&45&2.60&bri&44.0&fcc&2.69&0.38\\
  ~&90&2.60&top-hcp&89.7&fcc&3.67&0.60\\
  ~&90&2.60&bri&77.6&fcc&3.06&0.34\\
\hline
   ~&0&2.69&bri-fcc&21.6&fcc&2.93&0.46\\
bri-fcc&45&2.69&bri-fcc&39.5&hcp&2.86&0.40\\
  ~&90&2.69&bri-fcc&73.5&hcp&2.74&0.65\\
\hline
  ~&0&2.50&top-fcc&20.2&hcp&2.97&0.51\\
  ~&45&2.50&top-fcc&54.4&hcp&2.66&0.62\\
top&45&2.50&top-fcc&55.9&fcc&2.67&0.64\\
  ~&90&2.50&top-fcc&73.4&hcp&2.69&0.64\\
  ~&90&2.50&top-fcc&73.0&fcc&2.73&$\underline{0.66}$\\
\hline
\end{tabular}
\end{center}
\label{phys}
\end{table}
Table \ref{phys} shows that the adsorption energy for the stable non-dissociative structure is 0.66 eV, 
and the adsorption site preferred by the sulfur atom is located in the fcc region 
(at fcc but toward the atop site, top-fcc), in keeping with 
Refs. \cite{gca},\cite{rlm}. Kondoh et al. observed that methanethiolate occupies the atop site on the perfect Au(111) 
surface \cite{kis}, but here our adsorbate is methanethiol. From Table I, it leans toward the atop site.
The typical S-Au bond length is around 2.7 $\AA$, which indicates that non-dissociative 
adsorption may be understood as chemisorption \cite{sf},\cite{rlm}.

To compare dissociative and non-dissociative adsorption energies,  we include dissociative geometries 
and adsorption energies of methanethiolate on Au(111) at 0.25ML. Here the hydrogen atom is detached 
from the sulfur atom and forms a bond with the Au atom. As underlined in Table \ref{chem}, the maximum
adsorption energy is 0.06 eV. The favored region for methanethiolate is the fcc region which matches
the results reported in \cite{gca},\cite{yzr}-\cite{yr}.
\begin{table}
\caption{Dissociative geometries and adsorption energies for the configurations at 0.25 ML.}
\begin{center}
\begin{tabular}{cccccccc}
\hline
initial & $\theta$ &    $d_{S-Au}$ & optimized & $\theta$ & tilt     & $d_{S-Au}$ & $E_{ads}$\\
site    & ~        &  ~       & site      & ~        & direct & ~       & ~\\
\hline
  ~&0&2.60&fcc&2.7&fcc&2.45&-0.02\\
fcc&$\#$0&2.60&fcc&1.5&fcc&2.48&-0.09\\
  ~&45&2.60&bri-fcc&41.1&hcp&2.45&0.02\\
  ~&90&2.60&fcc&60.8&hcp&2.45&$\underline{0.06}$\\
\hline
  ~&0&2.60&bri-hcp&2.1&hcp&2.50&-0.14\\
hcp&45&2.60&bri-hcp&41.7&fcc&2.46&-0.02\\
  ~&90&2.60&bri-hcp&60.7&fcc&2.46&0.04\\
\hline
  ~&$\#$0&2.60&bri-fcc&19.8&hcp&2.84&0.40\\
bri&45   &2.60&bri-hcp&41.7&hcp&2.46&0.05\\
  ~&90&2.60&bri-hcp&60.7&fcc&2.46&0.03\\
\hline
   ~&0&2.69&fcc&4.4&hcp&2.46&-0.02\\
bri-fcc&45&2.69&bri-fcc&47.1&hcp&2.47&0.02\\
  ~&90&2.69&bri-fcc&63.1&hcp&2.46&0.04\\
\hline
  ~&$\#$0&2.50&top-fcc&38.3&hcp&2.69&0.57\\
top&45&2.50&top-hcp&58.3&fcc&2.38&-0.34\\
  ~&90&2.50&top&70.3&fcc&2.48&-0.37\\
\hline
\end{tabular}
\end{center}
\label{chem}
\end{table}
In Table \ref{chem} there are three configurations marked by $\#$. 
The first $\#$ structure for
methanethiolate itself is identical with the first one in the fcc rubric, except for a different 
location of the dissociated hydrogen atom. 
The initial structures of the second and third $\#$ configuration are dissociative, 
but in the course of the geometry optimization, the hydrogen atom reunites with the 
 sulfur atom. Thus the optimized structures are non-dissociative. 

Table \ref{phys} and Table \ref{chem} show that the adsorption 
energy of the non-dissociative structure is higher than that of the 
dissociative one by 0.6 eV. For further examination of our results we apply the ultrasoft pseudopotential
in conjunction with the PBE XC \cite{pbe} to
re-calculate the above structures, and find an adsorption energy difference of 0.58 eV between the cases of 
cleavage and non-cleavage. Since for the perfect Au(111) surface the non-dissociative structure 
displays a higher adsorption energy than the dissociative one by 0.6 eV, 
we conclude that at low or room temperature the adsorption of the 
methanethiol is non-dissociative, which supports the most recent experimental finding \cite{rlm}.

Next we turn to high coverage (1.0ML). We arrange four methanethiol molecules in the 
$c(4\times 2)$ supercell in keeping with the experimentally detected structure \cite{ccl},\cite{bsk}.
The molecules labeled
1(3) and 2(4) are symmetry equivalent, see Fig. \ref{4x2}a. The calculated results 
are shown in Table \ref{1ml}.
\begin{table}
\caption{The non-dissociative geometries and adsorption energies for the configurations at 1.0ML.}
\begin{center}
\begin{tabular}{cccccccc}
\hline
initial &$d_{S-S}$&$d_{S-Au}$ & optimized & $\theta$ & $d_{S-S}$& $d_{S-Au}$ & $E_{ads}$\\
site    & ~   &  ~       & site      & ~        &      ~  & ~       & ~\\
\hline
fcc+bri&2.24 &2.89&hcp+fcc &37.3&4.90&2.95&0.22\\
hcp+fcc&2.41 &2.50&hcp+hcp &43.1&3.76&2.64&0.19\\
bri+hcp&2.30 &2.45&fcc+fcc&30.0&3.89&3.28&0.19\\
top+fcc&2.30 &2.50&fcc+hcp&23.0&4.03&3.12&0.22\\
top+fcc$'$&2.20 &2.20&fcc+hcp&26.8&3.96&2.72&0.22\\
top+hcp&2.30 &2.50&bri+fcc &22.1&4.05&2.91&0.17\\
top+bri$\#$&2.54 &2.50&bri+fcc&38.5&2.38&2.93&-0.32\\
top+bri$'$&2.54 &2.50&fcc+hcp &14.3&4.13&3.23&0.20\\
fcc+fcc&5.08 &2.62&fcc+fcc&77.0&4.92&3.65&0.20\\
hcp+hcp&5.08 &2.50&hcp+hcp&73.2&5.08&3.54&0.22\\
fcc+hcp$\#$&3.39 &2.50&fcc+fcc&50.8&5.20&2.47&0.07\\
\hline
\end{tabular}
\end{center}
\label{1ml}
\end{table}
The notation fcc+bri in Table \ref{1ml} denotes that the sulfur atom of methanethiol 1
is placed at a fcc center and that of methanethiol 3 at 
bridge site, the notation hcp+fcc, top+fcc etc. is to be understood analogously. 
The configurations marked by
 $\#$ are characterized by non-dissociative initial structures
but dissociative equilibrium structures, as emerging from geometry optimization. Their adsorption energies
(-0.32 eV and 0.07 eV) are distinctly lower than those of the alternative intact structures (around 0.20 eV), which further confirms
our conclusion obtained for the low coverage case: the S-H bond is not broken when
the methanethiol molecules interact with a perfect Au(111) surface. 
At 1.0ML, the range of adsorption 
energies at various sites is within 0.05 eV, corresponding to an almost iso-energetic situation.
These calculated results
explain that the STM tip easily induces the motion of the
methanethiol molecules on the surface at high coverage \cite{rlm}. 
Table \ref{1ml} shows that some distances between two S atoms are
set to be less than 2.6 $\AA$, but after optimization, the distances become
larger than 3.75 $\AA$. 
For the seventh configuration in Table \ref{1ml}, however, $d_{S-S}$ is 2.38 $\AA$.
Here two of the hydrogen atoms have been detached from the sulfur atoms and
the adsorption energy is lower. Thus we conclude 
that for intact methanethiol molecules it is impossible for two 
sulfur atoms to form a dimer at high coverage, which is consistent with 
the observation reported in Refs \cite{kcm},\cite{vgs} and \cite{kis}. 

To explore the mechanism for S-H bond breaking in adsorbed methanethiol, we consider an Au(111) surface
with defects, and for simplicity, we assume a vacancy in the top layer of the $c(4\times 2)$ supercell. 
The calculation shows that the formation energy of the vacancy on Au(111) is 0.6 eV, and the 
introduction frequency of the vacancy 
is $1.5\times 10^{14}$ Hz  \cite{mh}.
The geometries and adsorption energies for the optimized 
non-dissociative and dissociative configurations at 0.25 ML are indicated  in Table \ref{defect}.
\begin{table}
\caption{The geometries and adsorption energies for the non-dissociative and
dissociative configurations at 0.25 ML. The nomenclature $above$ or $below$ refers to positions of the S or the H atom above or below
the vacancy. The terms $one$ $bond$ or $two$ $bonds$ denote S atom 
bonding to one or two gold atoms. The term $embedded$ means that the vacancy is filled by the S atom.}
\begin{center}
\begin{tabular}{ccccccccc}
\hline
initial &$d_{S-Au}$ & optimized         & optimized & $d_{S-Au}$ & $E_{ads}$\\
site (S)&  ~        & site (S)          &  site (H) & ~          & ~\\
\hline
&       &    &  not-cleaved &                       &           &            \\
 above &3.86& above&-&3.55&0.54\\
 embedded&2.91& above&-&3.14&0.44\\
 one bond&2.50& above &-&2.62&$\underline{0.73}$\\
\hline
&       &     &  cleaved &                  &           &            \\
above &3.86& two bonds &above Au&2.44&0.55\\
above &3.86& two bonds&bridge 2 Au &2.44&$\underline{0.81}$\\
embedded &2.91& one bond &above Au&2.54&0.17\\
 one bond&2.50& two bonds&below &2.42&0.37\\
 one bond&2.50& one bond&above Au &2.36&-0.14\\
one bond&2.50& one bond&bridge 2 Au &2.37&0.15\\
\hline
\end{tabular}
\end{center}
\label{defect}
\end{table}
It is shown that in the presence of the vacancy, the adsorption energy of the
stable dissociative configuration (0.81 eV) is higher than that of the non-dissociative configuration (0.73 eV),
in exact opposition to the vacancy-free case. The vacant site is located at the center of the gold atom hexagon. 
The most stable structure is realized when the sulfur and hydrogen
atoms attach to two different gold atom pairs within the same 
hexagon, see Fig. \ref{4x2}b. If one attaches the sulfur atom on the other gold atoms instead of the six gold atoms surrounding the 
vacancy, 
the adsorption energy is almost equal to that found for the vacancy-free situation, since the vacancy
exerts little effect on the S atom. 
The essential conclusion is that the defected Au(111) acts as a catalyst for the S-H bond rupture 
while the perfect Au(111) surface does not. It is expected that with increasing density of defects,
the S-H bond scission becomes the prevailing mode of interaction between the methanethiol adsorbate and 
the Au(111) substrate. Here we point out that after the loss of the H atom, 
the strength of the bond between the S atom and the defected Au surface exceeds that 
between the S atom and the perfect Au surface, the difference amounting to about 0.75 eV. 
This is consistent with the result reported in Ref. \cite{mh}.

\begin{figure}
\includegraphics[width=3.5in]{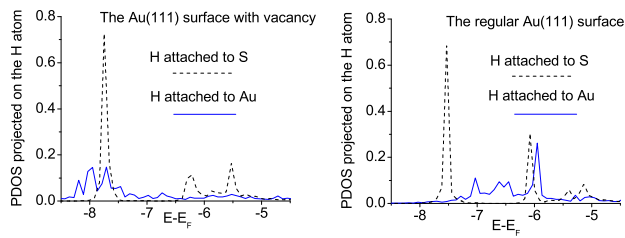}
\caption{The PDOS projected on the H atom initially attached to the S atom adsorbed on defected and 
perfect Au(111).}  
\label{hpdos}
\end{figure}

To explore this phenomenon in greater detail,
we have calculated the partial density of states (PDOS) for the H atom initially attached to 
the S atom, see Fig.\ref{hpdos}. 
For the perfect surface, the primary peaks of the hydrogen PDOS for the non-dissociative 
structure are at positions of lower energy than those for the dissociative situation. The 
electronic configuration related to the former case is thus more stable than that corresponding 
to the latter, and the hydrogen atom forms a stronger bond with the sulfur atom than with the gold surface. 
For the defected surface, this order of stabilities is reversed. The primary peaks of the 
hydrogen PDOS for the dissociative structure turn out to be at lower energy than those of 
the non-dissociative one, and therefore the hydrogen atom forms a stronger bond with the 
gold surface than with the sulfur atom. 
For the hydrogen PDOS related to the non-dissociative structure on the defected surface, 
the highest peak exceeds the second highest very substantially. This distribution resembles 
more the DOS of the isolated hydrogen than the analogous spectrum for the perfect surface, 
indicating that the H-S bond is weak on the defected surface.

In summary, we predict for the first time that the S-H bond remains intact when methanethiol 
is adsorbed on the regular Au(111) surface, i.e., non-dissociative
adsorption is thermodynamically stable. However, the bond breaks as admission is 
made for defects in the Au(111) surface. At low coverage, 
the fcc region is favored for S atom adsorption. At saturated coverage, the 
various adsorption sites turn out to be almost iso-energetic, which explains satisfactorily that the STM 
tip easily induces the motion of 
methanethiol molecules on the surface at high coverage \cite{rlm}. Our calculation indicates that two 
intact methanethiol molecules do not form a dimer on the perfect Au(111) surface, 
which is consistent with the respective observation in Refs. \cite{kis}, \cite{kcm} and \cite{vgs}. 
These findings resolve the controversial issue 
of bond cleavage at low or room temperature. The emergence of thiolate on the Au(111) surface reflects 
the presence of surface
defects. Defects might be caused in the process of gold surface production,
or temperature enhancement \cite{kcm} or exposure to X-rays \cite{jsm}.
The sulfur dimers were proposed on the basis of X-ray diffraction (XRD) data obtained at saturated
coverage \cite{fee}, but we have found that a pair of intact methanethiols cannot form a dimer on the defect-free Au(111) 
surface. We are in the process of exploring the conditions for dimerization, and in particular inspect defect and 
temperature related effects.

We thank Dr. W. Hase for a helpful discussion. 
This work is supported by the NSF through the grants HRD-9805465 and DMR-0304036, 
by the NIH through the grant S06-GM008047, and by the AHPCRC under Cooperative Agreement 
No. DAAD 19-01-2-0014. 
   
\end{document}